# Model predictive control strategy in waked wind farms for optimal fatigue loads

Cheng ZHONG[a,b], Yicheng DING[b], Husai WANG[b], Jikai CHEN[a,b], Jian WANG[a,b], Yang LI[a,b]

[a]*Key Laboratory of Modern Power System Simulation and Control &Renewable Energy Technology Ministry of Education (Northeast Electric Power University), Jilin, Jilin Province 132012, China*
[b]*Northeast Electric Power University College of Electrical Engineering, Jilin, Jilin Province 132012, China*

**Abstract**

With the rapid growth of wind power penetration, wind farms (WFs) are required to implement frequency regulation that active power control to track a given power reference. Due to the wake interaction of the wind turbines (WTs), there is more than one solution to distributing power reference among the operating WTs, which can be exploited as an optimization problem for the second goal, such as fatigue load alleviation. In this paper, a closed-loop model predictive controller is developed that minimizes the wind farm tracking errors, the dynamical fatigue load, and and the load equalization. The controller is evaluated in a medium-fidelity model. A 64 WTs simulation case study is used to demonstrate the control performance for different penalty factor settings. The results indicated the WF can alleviate dynamical fatigue load and have no significant impact on power tracking. However, the uneven load distribution in the wind turbine system poses challenges for maintenance. By adding a trade-off between the load equalization and dynamical fatigue load, the load differences between WTs are significantly reduced, while the dynamical fatigue load slightly increases when selecting a proper penalty factor.

*Keywords:* model predictive control, wind farm control, fatigue load, load equalization, power tracking

## 1. Introduction

The development trend of clean energy is irreversible [1], and the use of wind power has rapidly increased in the past ten years, playing a leading role among renewable energy sources. Because wind turbines (WTs) are placed together on a farm, the wakes of upstream turbines influence the performance of downstream turbines. The control of a WF is challenging because of aerodynamic interactions. This wake effect reduces the wind speed and increases turbulence [2]. The former diminishes the captured wind power [3][4], while the latter increases the fatigue load of the turbine [5]. As the penetration level of wind power increases, wind farms (WFs) must adhere to more stringent technical requirements, such as regulating their active power to support the grid frequency while simultaneously reducing the fatigue load on the entire WF [6].

The development of frequency control in WFs is relatively advanced. A common method is virtual inertia control, which reduces the rate of change of frequency (RoCoF) and frequency nadir by temporarily injecting power [7]. Conversely, primary frequency control reduces frequency deviations by adjusting the active power [8]. More advanced methods are also available, such as renewable energy load-frequency control based on demand response [9]. This study primarily focuses on optimizing the fatigue load during frequency regulation. Research on additional frequency control strategies is limited; therefore, only the virtual inertia and primary frequency control are employed.

During frequency control, WTs must actively adjust power outputs to respond to power demands, which leads to increased fatigue loads on WFs [10]. To achieve power reference tracking on a WF, where the given power reference is lower than the maximum optimal power, the total output power and wake effects must be considered when allocating power reference values to each individual WT. Because multiple solutions can be used to achieve the same power reference, it is necessary to introduce additional performance metrics in the WF power tracking problem, such as the derivative of the thrust coefficient [13][14], dynamic fatigue loads [18-20], and load equalization [22][23][32].

Model predictive control (MPC) has been widely used in recent research to address the WFs optimization problem of active power control. In [11], the authors proposed a distributed MPC controller that incorporated a simplified WF model to evaluate the performance of the controller for power tracking. Reference [12] considered the stochastic nature of a power reference signal; however, the optimization objective was solely power tracking. In [13], a closed-loop MPC was employed, with the weighted sum of the tracking error and changes in the thrust coefficient as the optimization objective. Reference [14] incorporated both deviations in the thrust coefficient and the derivative of the thrust coefficient into the objective to prevent high-frequency oscillations, and the proposed controller was tested using a high-fidelity model. The output of WFs can be improved by adjusting the thrust coefficients [15][30] and yaw angles [16][17]. These adjustments significantly enhance the optimization of power tracking. However, the control strategies proposed in previous studies mainly focused on the power output of the WF and did not address the fatigue loads generated by WTs. And [33] indicated that wind-induced fatigue shortens fatigue life as the yaw angle increases, yaw conditions should be avoided during wind turbine operations.

As the WTs track a given power reference, the axial force varies with time, increasing the fatigue load on the WT. Reference [18] presented a distributed-model predictive control method for minimizing the load variation to reduce the dynamic fatigue load. In [19], the authors considered the stochasticity of wind speed in the MPC controller and took the load variation as one of the optimization objectives. In [20], the thrust coefficients that provided WF power tracking while minimizing the dynamic fatigue load were evaluated with constrained MPC and yaw settings that increased the available WF power. However, even if the load variation is minimized, the upstream turbines may suffer from high loads that lead to excessive wear. In [21], the authors proposed a load-equalization concept to balance the lifetimes of individual WTs. In [31], by equalizing the structural load distribution across the turbines, the degradation of highly loaded turbines and their associated maintenance costs are reduced. Reference [22] presented a closed-loop wind-farm active power control framework that equalizes the structural loads of WTs. In [23], structural load equalization was used as an optimization objective for active power control, which is beneficial for having all the WTs similarly loaded or worn. In [32], the authors proposed a model predictive control strategy to solve the problem of uneven fatigue distribution in offshore wind farms by adjusting the reference value of wind turbine active power, however, the effect of dynamic fatigue loads was not taken into account.

Some of the previous articles have reduced fatigue loads by optimizing the load distribution of WTs, but as mentioned before, optimizing only dynamic fatigue loads may lead to high fatigue loads in some WTs. Owing to the solution flexibility of the MPC controller, in this study, a closed-loop MPC controller is proposed to achieve power tracking, dynamic fatigue load reduction, and load equalization. We found the controller can effectively address uneven fatigue distribution, and maintain the impact on dynamic fatigue loads and power-tracking error within a lower range. A medium-fidelity WF model, WindFarmSimulator (WFsim), was used to validate the effectiveness of the proposed scheme [24].

The remainder of this paper is organized as follows: Section II introduces the employed WF model Section III describes the proposed MPC controller, Section IV presents the simulation results, and Section V concludes the paper.

## 2. WF model

### *2.1. Flow model*

This study used the incompressible two-dimensional (2D) Navier-Stokes equations, referred to as WFsim modeling [24], which are constrained by the continuity equations:

$$\frac{\partial u}{\partial t}+\left(\frac{\partial u}{\partial x},\frac{\partial u}{\partial y}\right)\cdot\vec{U}=f_x-\frac{1}{\rho}\frac{\partial p}{\partial x}-\frac{\partial \tau_H}{\partial x}$$
$$\frac{\partial v}{\partial t}+\left(\frac{\partial v}{\partial x},\frac{\partial v}{\partial y}\right)\cdot\vec{U}=f_y-\frac{1}{\rho}\frac{\partial p}{\partial y}-\frac{\partial \tau_H}{\partial y}\ , \tag{1}$$
$$\frac{\partial u}{\partial x}+\frac{\partial v}{\partial y}=-\frac{\partial v}{\partial y}$$

where $\vec{U}=(u,v)^T$ is the velocity vector field at the hub height of the WT; $u$ and $v$ are the wind speeds in the x- and y- directions, i.e., longitudinal and lateral directions, respectively; $\rho$ denotes the air density, $p$ denotes the pressure field; and $\tau_H$ is a 2D tensor, which is modeled in detail in [24]. $f_x$ and $f_y$ are the external force terms in the x- and y- directions, respectively.

Because the study does not consider yaw angle control, to simplify the calculation, lateral thrust force $f_y$ is neglected. The 2D flow is spatially and temporally discretized over a specified staggered grid following [25]; therefore, the turbine model is incorporated into the flow model using discretized external forces, and the longitudinal thrust forces $f_x$ can be expressed as [30]

$$f_x=\frac{1}{2}\rho\Delta y C_T^{'}U^2\ , \tag{2}$$

where, $\Delta y$ is the cell grid width. $U=\sqrt{u^2+v^2}$ is the wind speed at the plane of the blade disc. $C_T^{'}$ is the local thrust coefficient is obtained from the individual wind turbine model in section 3.1.

### *2.2. Single-machine WT model*

During the wind-energy conversion process, the rotor of the WT is subjected to a secondary effect that dissipates a portion of the wind energy. This effect is manifested as a force perpendicular to the rotor plane, resulting in a bending moment on the tower and subsequent oscillations of the WT. This force is known as the thrust force, and the thrust of an individual WT due to the flow is defined as [20]

$$F=\frac{1}{2}\rho A C_T^{'}U^2\ . \tag{3}$$

In the initial stage of wind-energy conversion, we employed an aerodynamic model of the rotor to characterize the process. Our focus lies on modeling the conversion of the energy harnessed by the rotor into the driving torque of the rotating machine, ultimately resulting in the generation of power by the WT.

$$P=\frac{1}{2}\rho A C_T^{'}U^3\ , \tag{4}$$

where, $A$ is the area of the WT.

## 3. Control strategy

This study aimed to establish frequency control and the optimal fatigue load for a WF. The control diagram is shown in Fig. 1.

The WF control system is divided into two layers: central and individual WT controls. The central controller consists of frequency-control and closed-loop MPC blocks. The frequency-control block receives the power dispatch command or grid frequency measurement and then generates the total power reference of the WF. The frequency-control block includes virtual inertia and a droop controller [28].

The total power reference of the WF is [29]

$$P_{ref}=P_{command}+K_D\Delta f+K_I\frac{df_{measure}}{dt}\ , \tag{5}$$

where $P_{command}$ is the power command; $P_{ref}$ is the reference power with the addition of virtual inertia and droop control; $f_0$ is the reference frequency; $f_{measure}$ is the measured frequency; $K_D$ is the droop coefficient; and $K_I$ is the virtual inertia coefficient.

The closed-loop MPC blocks receive $P_{ref}$, wind speed $U$, and the feedback $\left[F \quad P \quad C_T'\right]^T$ from the WF monitoring system. Then, it solves the optimal real-time problem to obtain single-machine power commands for each WT, that is $P_i^*$. These power commands are dispatched to individuals via communication lines. An individual WT regulates the rotor speed and pitch angle to track the power commands and feed the operating states back to the central controller.

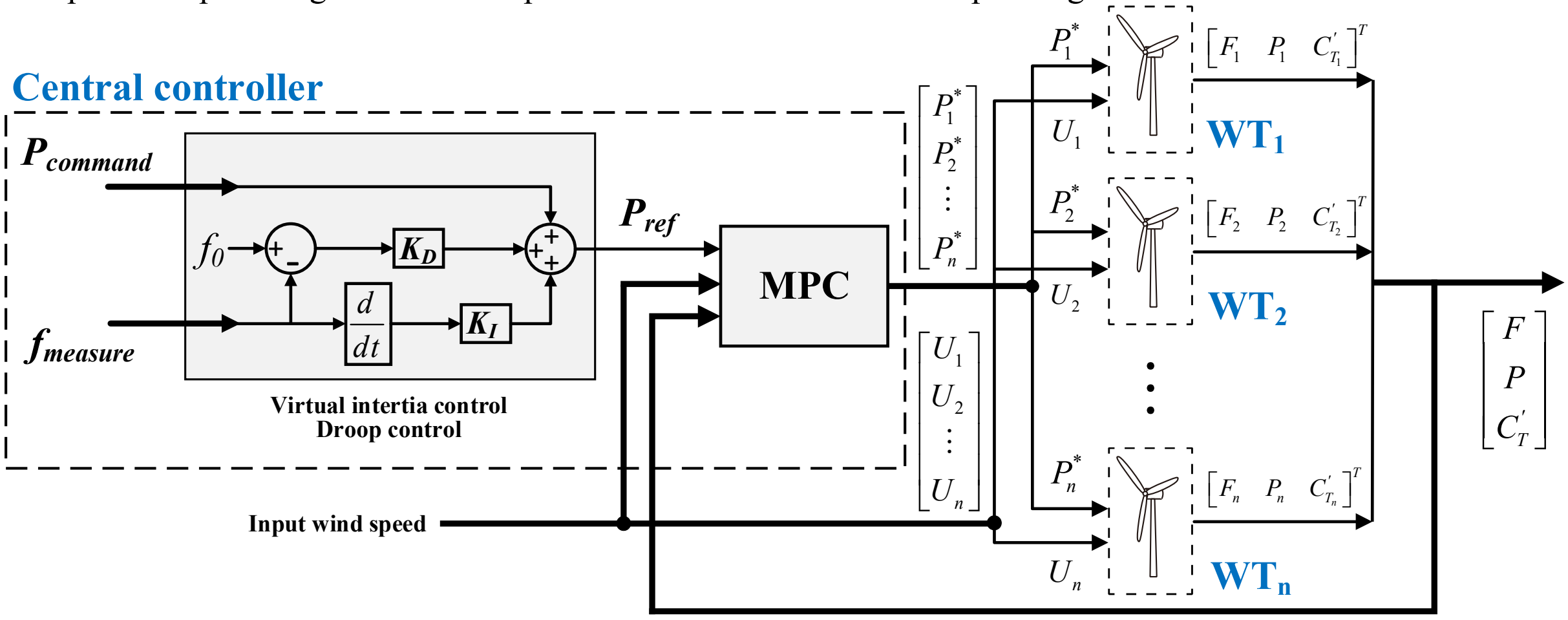

Fig. 1. Control structure of the proposed closed-loop MPC for WFs.

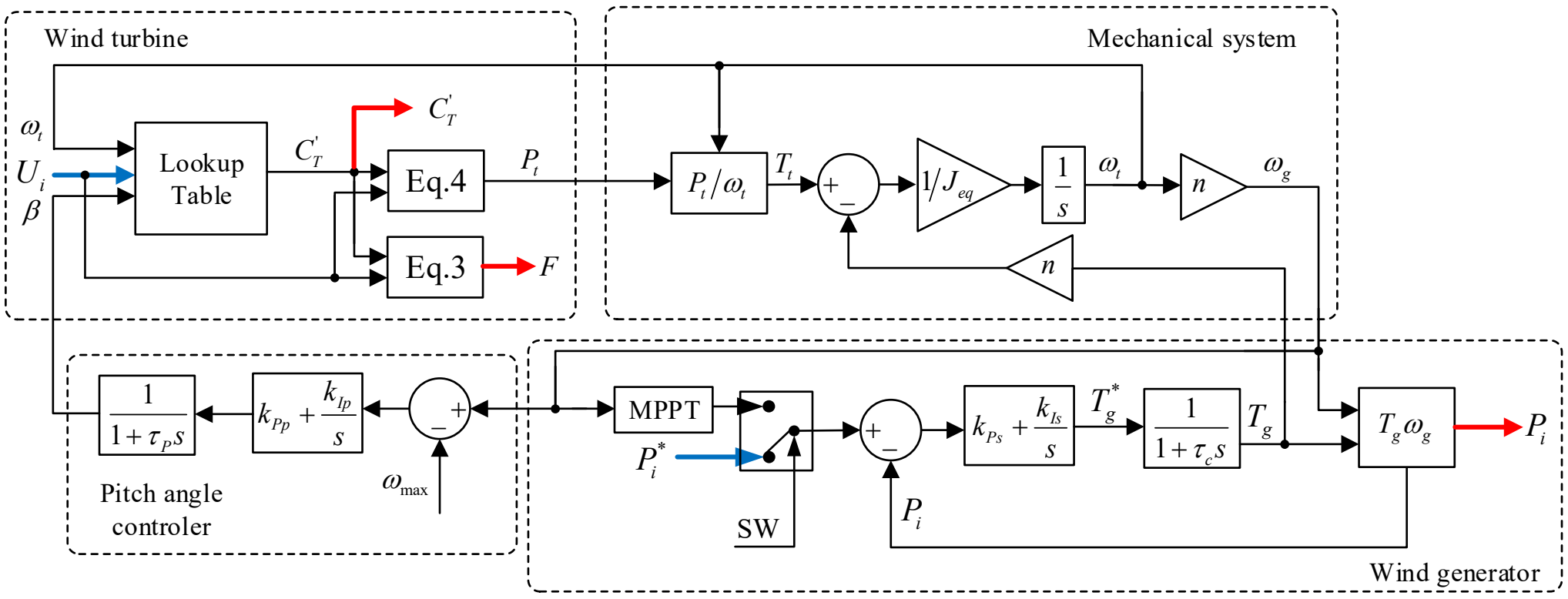

Fig. 2. WT control system.

## 3.1. Individual WT active power control

Individual WT modeling and control systems are shown in Fig. 2, as described in [26]. $T$, $P$, and $\omega$ represent torque, power, and angular speed, respectively; subscripts $g$ and $t$ are used to indicate the generator and turbine variables, respectively; $\tau$ is the time constant; $k$ is the PI controller coefficient; and $n$ is the gear-box ratio. The dynamics of the WT are represented as a single-mass model, and the equivalent rotational inertia is denoted as $J_{eq}$.

The reference power $P_i^*$ can be obtained from the Maximum Power Point Tracking (MPPT) or central control command mode, which is determined by switch SW. The electrical torque $T_g$ is obtained by a PI control and a first-order inertia element, where the PI control is the speed controller and the first-order inertial element represents the control delay of the converter. The rotor speed $\omega_t$ is adjusted to track the reference power. A pitch-control loop is used to ensure that the WT does not exceed its rotor limit or power rating. In Fig. 2, $C_T'$ is obtained by a two-dimensional look-up table method. In this table the horizontal coordinate is the tip speed ratio $\lambda = \frac{w_t r}{U_i}$ ($r$ is the radius) and the vertical coordinate is the pitch angle $\beta$.

### *3.2. MPC controller*

In this study, the MPC was used to minimize the power-tracking error and fatigue load and to obtain the dispatching power command for each WT. The MPC is a closed-loop optimal-control method based on an internal model in a finite time horizon that satisfies the imposed constraints and inputs. An MPC is typically composed of three parts: a prediction model, feedback correction, and rolling optimization. The MPC controller takes the current state as the initial state. The system states in the prediction horizon were predicted by the controls at the present or past sampling times. The optimal control variables were obtained by solving a finite-time optimization problem using real-time rolling. Subsequently, only the first term of the optimal control variables was used in the actual system.

The control diagram of the proposed method is shown in Fig. 3.

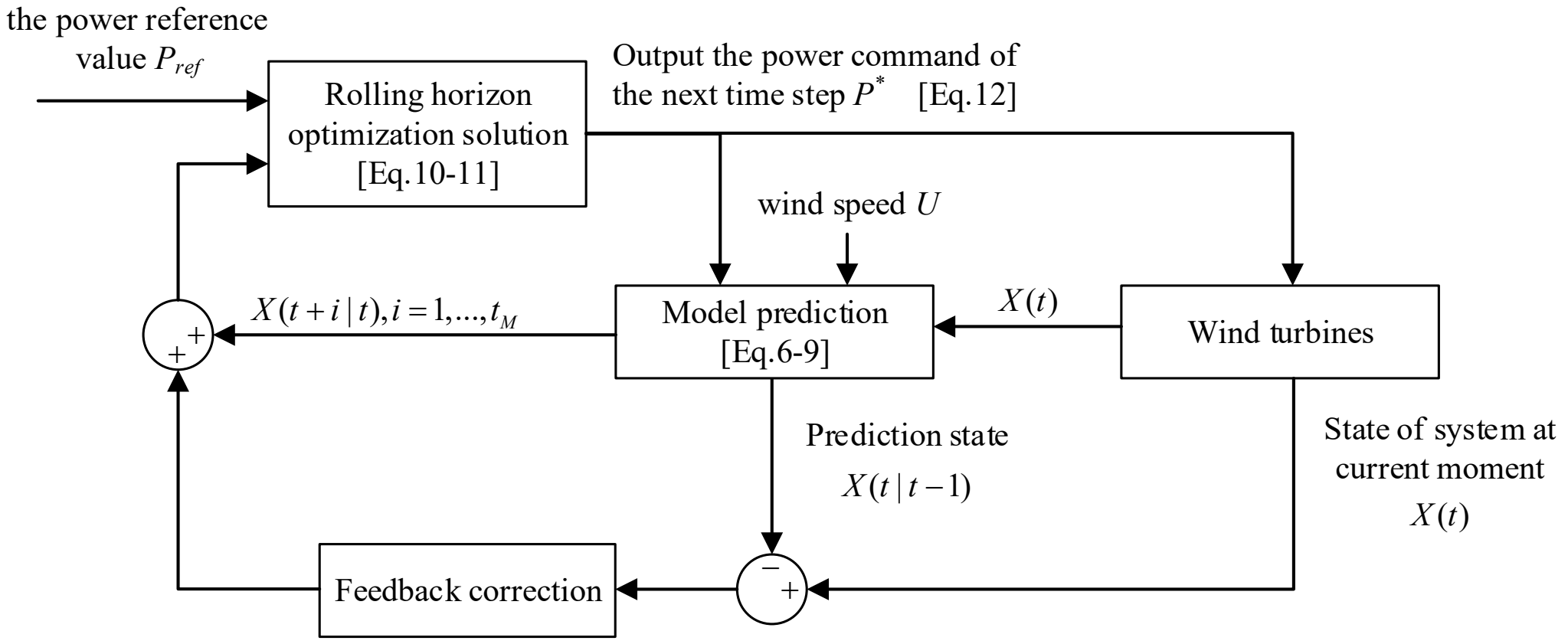

Fig. 3. Schematic of the MPC controller.

#### *3.2.1. Prediction model*

In the MPC scheme, a prediction model is used to predict the states of WTs. The dynamic model of the WT in the MPC controller is described as [20]

$$
\begin{aligned}
F_i(t) &= \frac{1}{2}\rho A \hat{C}_{T_i}'(t)[U_i(t)]^2 \\
P_i(t) &= \frac{1}{2}\rho A \hat{C}_{T_i}'(t)[U_i(t)]^3 \\
C_{T_i}'(t) &= \tau \frac{d\hat{C}_{T_i}'(t)}{dt} + \hat{C}_{T_i}'(t)
\end{aligned}
\tag{6}
$$

where $F_i(t)$ and $P_i(t)$ denote the axial force and power output of the $i^{th}$ WT at sampling time $t$, respectively. $\hat{C}_{T_i}'(t)$ is the first-order filter signal of $C_{T_i}'(t)$, and $\tau$ is the time constant of the filter. A first-order low-pass filter was used to smoothen the control signals. A zero-order keeper was used to discretize the control variables.

Finally, the state space expression of $N$ WTs was obtained as

$$X(t+1)=\begin{bmatrix} A_1 & & & \\ & A_2 & & \\ & & \ddots & \\ & & & A_N \end{bmatrix} X(t)+\begin{bmatrix} B_1(t) & & & \\ & B_2(t) & & \\ & & \ddots & \\ & & & B_N(t) \end{bmatrix}\begin{bmatrix} \hat{C}'_{T_1}(t) \\ \hat{C}'_{T_2}(t) \\ \vdots \\ \hat{C}'_{T_N}(t) \end{bmatrix} \tag{7}$$

which is simplified as

$$X(t+1)=AX(t)+B\hat{C}'_T(t) \tag{8}$$

with

$$X(t)=\begin{bmatrix} X_1(t) \\ X_2(t) \\ \vdots \\ X_N(t) \end{bmatrix}\in\mathbb{R}^{3N},\ X_i(t)=\begin{bmatrix} F_i(t) \\ P_i(t) \\ \hat{C}'_{T_i}(t) \end{bmatrix}\in\mathbb{R}^3,\ A_i\in\mathbb{R}^{3\times 3},\ B_i(t)\in\mathbb{R}^3$$

The wind speed in the WT was not significantly affected by the wake [25]. The wake effect is neglected in the surrogate dynamic model, and only the turbine dynamics are considered in this study. In other words, wind speed was assumed to remain unchanged during the horizon prediction period.

Hence, according to the state–space equation, Equation (8), the entire WF prediction model for future M-steps was established.

$$\begin{aligned}\begin{bmatrix} X(t_0+1) \\ X(t_0+2) \\ X(t_0+3) \\ \vdots \\ X(t_0+M) \end{bmatrix} &= \begin{bmatrix} A & & & & \\ & A^2 & & & \\ & & A^3 & & \\ & & & \ddots & \\ & & & & A^M \end{bmatrix}\begin{bmatrix} X(t_0) \\ X(t_0) \\ X(t_0) \\ \vdots \\ X(t_0) \end{bmatrix} \\ &+\begin{bmatrix} B & & & & \\ AB & B & & & \\ A^2B & AB & B & & \\ \vdots & \vdots & \vdots & \ddots & \\ A^{M-1}B & A^{M-2}B & A^{M-3}B & \cdots & B \end{bmatrix}\begin{bmatrix} \hat{C}'_T(t_0) \\ \hat{C}'_T(t_0+1) \\ \hat{C}'_T(t_0+2) \\ \vdots \\ \hat{C}'_T(t_0+M) \end{bmatrix}+\begin{bmatrix} \mu(X(t_0)-X(t_0\,|\,t_0-1)) \\ 0 \\ 0 \\ \vdots \\ 0 \end{bmatrix}\end{aligned} \tag{9}$$

where $\mu=dig(\mu_1,\mu_2,\mu_3,\ldots,\mu_1,\mu_2,\mu_3)\in\mathbb{R}^{3N\times 3N}$ is the correction matrix used to improve the accuracy of the prediction model.

### *3.2.2. Rolling optimization*

During the operation of a WT, it is necessary to ensure accurate tracking of the power reference, minimize the dynamic fatigue load, and simultaneously reduce the load deviation from its average value. Hence, the objective function is defined as follows:

$$\begin{aligned} J=\min\sum_{t=t_0}^{t=t_0+M} & \Delta P_t^T Q\Delta P_t+(C'_T(t)-C'_T(t-1))^T R(C'_T(t)-C'_T(t-1)) \\ & +w\cdot[(F(t)-F(t-1))^T S_1(F(t)-F(t-1))+(F(t)-F_{mean}(t))^T S_2(F(t)-F_{mean}(t))] \end{aligned} \tag{10}$$

with $S_1=s\cdot I_{(N\cdot M)\times(N\cdot M)}$, $S_2=(1-s)\cdot 10^{-2}\cdot I_{(N\cdot M)\times(N\cdot M)}$, $\Delta P_t=P_{ref}(t)-\sum_{i=1}^{N}P_i(t)$

There are four terms in the object function. The first term represents the power reference-tracking goal, where $\Delta P_t$ is the error between the power output and reference power signal, the second term is the penalty for the variation of the power output to prevent high-frequency oscillations, the third term is the penalty for the load variation, and the fourth term is the penalty for the load deviation from its average value, multiplied by $10^{-2}$ to keep them in the same order of magnitude. $F(t)$ is the axial force of all WTs, and $F_{mean}(t)$ is the mean value of the axial force of all WTs at sampling time *t*. *Q, R, and S* are the weight matrices of the power-tracking error, local thrust coefficient, and load distribution, respectively. $Q$ is $1$, which is a *M*-dimensional unit matrix, *R* is defined as $r \cdot I_{(N \cdot M)\times(N \cdot M)}$. In this study, *r* was set to a constant value. We adjusted penalty factor *w* to trade-off the weight between the power tracking and load distribution, and adjusted penalty factor *s* to trade-off the weight between the load variation and load equalization.

The following constraints are imposed on $C'_{T_i}(t_i)$

$$\begin{aligned} & C'_{T\min} \le C'_{T_i}(t_i) \le C'_{T\max} \\ & \left| C'_{T_i}(t_i) - C'_{T_i}(t_i - 1) \right| \le dC'_T \end{aligned} \tag{11}$$

where $C'_{T\min}$ and $C'_{T\max}$ represent the upper and lower limits of $C'_T$, respectively, for each WT; $dC'_T$ represents the value of the permissible $C'_T$ variation. The objective of this rolling optimization problem is Eq. (10) and the constraint is Eq. (11), which is a typical nonlinear quadratic optimization problem. CPLEX is a high-performance commercial mathematical optimization software package developed by IBM. In this study, CPLEX was used to solve this real-time nonlinear quadratic programming problem and obtain the optimal control variable matrix of the prediction horizon. The optimal control variables at $t_0$+1 are used and converted into a power signal $P^*$ dispatched to the WTs, as shown in (12).

$$P^* = \begin{bmatrix} P_1(t_0+1) \\ P_2(t_0+1) \\ \vdots \\ P_N(t_0+1) \end{bmatrix} \tag{12}$$

## 4. Simulation and results

In this study, a large WF comprising 64 WTs was considered to validate the proposed control method. The WF layout and wake dynamic trends are shown in Fig. 4. The aerodynamic interactions were simulated using the WFsim toolbox, which models the mutual effects among turbines. In this study, two wind-farm scenarios were set up to test the applicability of the controller. Fig. 5a and 5b show the input wind speed conditions of the WF with average wind speeds of 9 and 12 m/s, respectively. Random perturbations with turbulence intensity $\sigma = 0.1$ were added to the wind speeds. Different power reference signals were set for each wind speed condition; Fig. 5c and 5d show the power reference signals for wind speeds of 9 and 12 m/s, respectively. The red solid curve represents the dispatching power $P_{command}$ for the grid operator, and the blue solid curve represents the total power reference $P_{ref}$ from the virtual inertia loop and droop control, which is the power-tracking reference in the MPC controller. The frequency-measured data used in the control system were obtained from a phasor-measurement unit (PMU), as depicted in Fig. 5e.

The aerodynamic model is composed of ambient field and wake flow models described in Section II. The ambient field was modeled as the hub height, and turbulent wind flowed under the assumption of Taylor turbulence. The WT model described in Section III was used. The proposed closed MPC control method was linked to the WFsim tool. The simulation duration was 900 s after the WF operation stabilization. The sampling period was 1 s, and the prediction horizon *M* was 10. $C'_{T\min}$ = 0.1, which means that the WT was not allowed to stall, and $C'_{T\max}$ = 2, which is the Betz-optimal value of the maximum extractable wind energy. $dC'_T$ was set to 0.2.

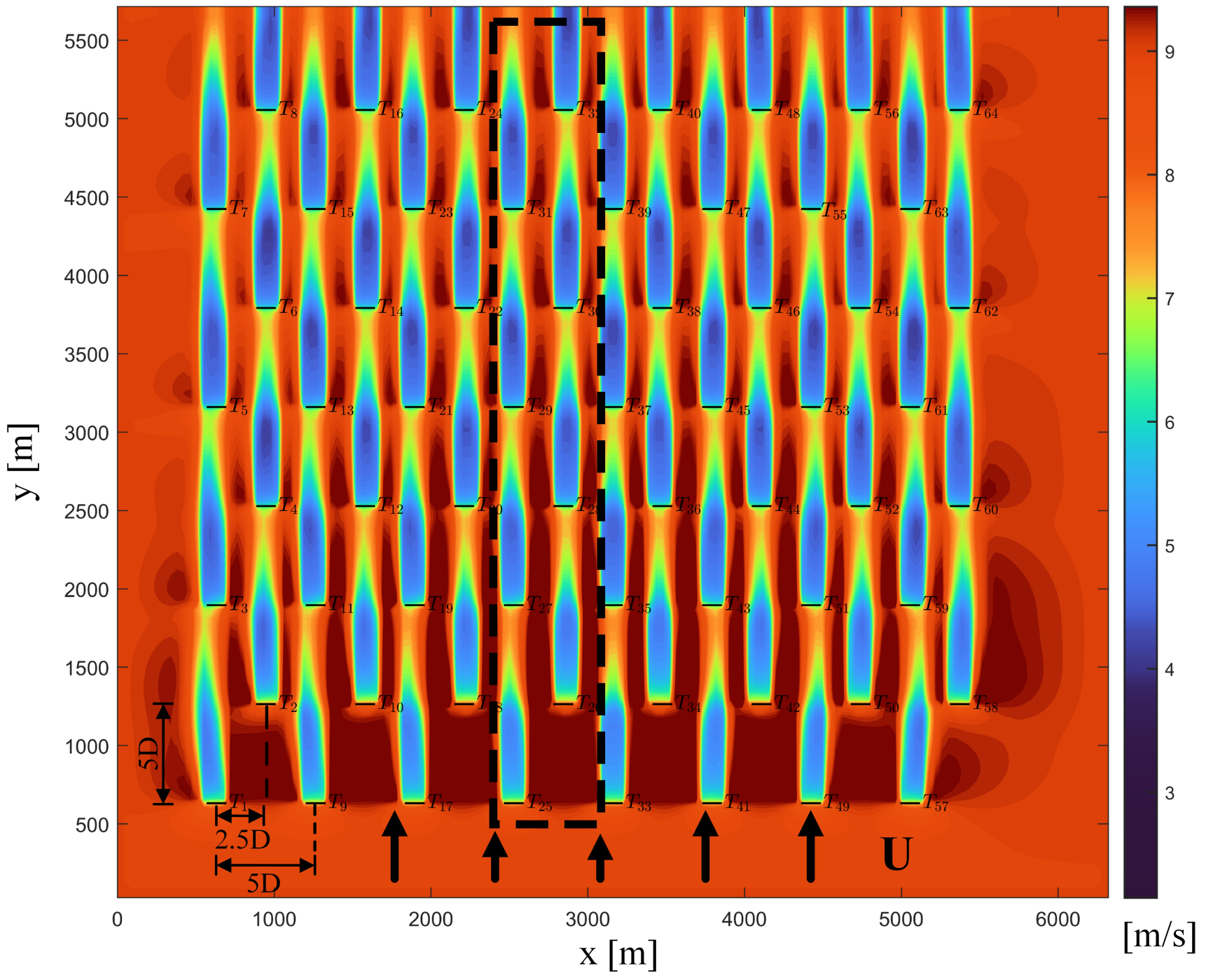

Fig. 4. The flow is moving from south to north, and the black vertical lines represent the wind turbines (WTs).

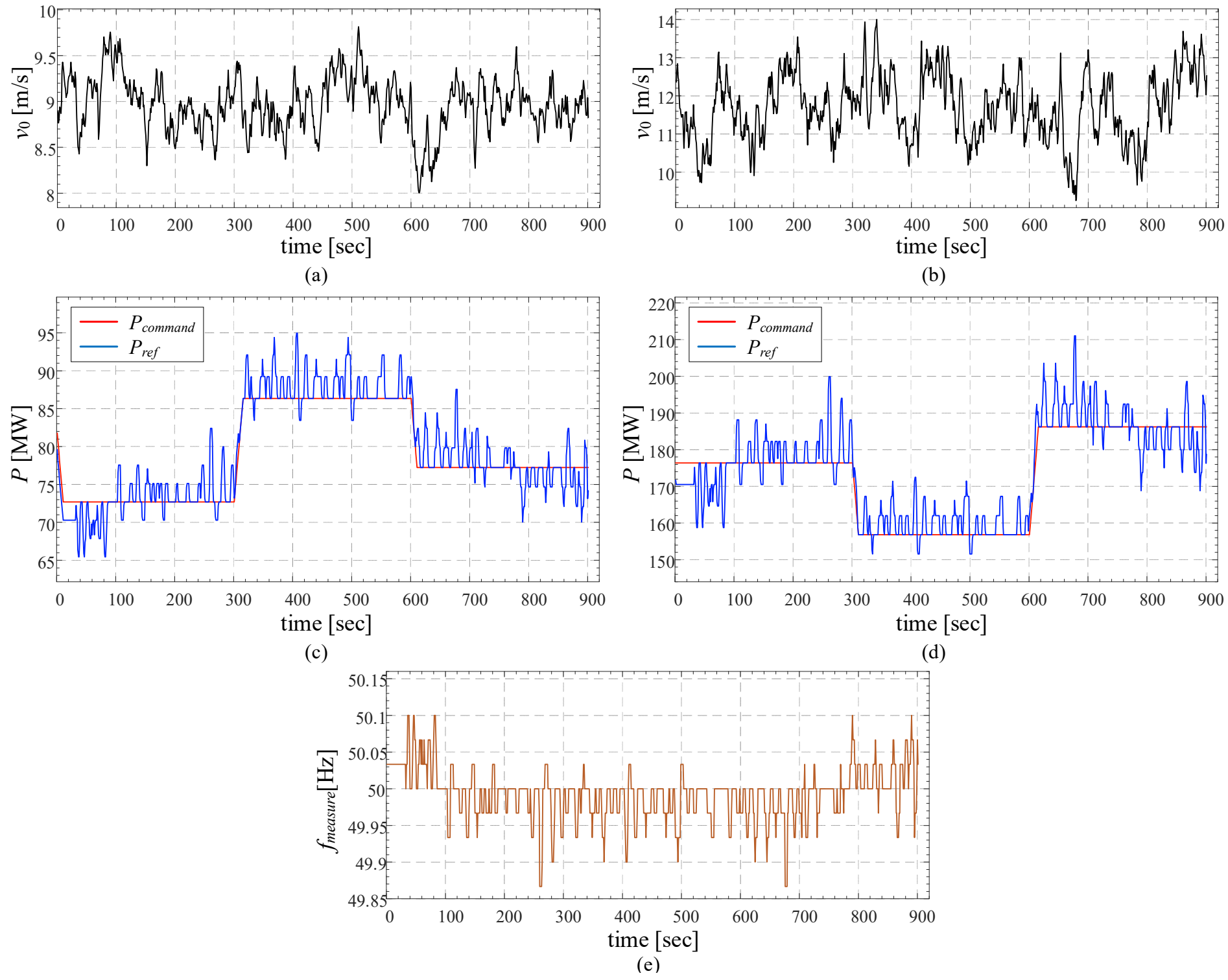

Fig. 5. Wind speed, power reference, and frequency curves.

A recently developed MPC algorithm [20] that assigns thrust coefficients to WTs to achieve power tracking and reduce the dynamic load on the WF was used for comparison. The optimization objective of this algorithm is given by Equation (13).

$$J = \min \sum_{t=t_0}^{t=t_0+M} (P_{ref}(t) - \sum_{i=1}^{N} P_i(t))^T Q (P_{ref}(t) - \sum_{i=1}^{N} P_i(t)) + (F(t) - F(t-1))^T W (F(t) - F(t-1)) . \quad (13)$$

Optimization objective (13) only reduces the dynamic fatigue loads without considering load equalization. In other words, it is represented by setting penalty factor $s = 1$ in Equation (10). The effectiveness of the proposed control strategy was verified by conducting comparative tests under the same conditions.

*4.1. Average wind speed of 9 m/s.*

Fig. 6a shows the power-tracking curve obtained using the control strategy from [20]. Penalty factor $w$ was used for the trade-off between the power-tracking error and dynamic fatigue load. The figure shows power-tracking curves for $w$ = 0, $1\times10^2$, $1\times10^3$, and $1\times10^4$ with $s$ = 1. The root mean square (RMS) errors are indicated in the legends. A large power-tracking error occurred when $w$ was set to $1\times10^4$. When $w$ was below $1\times10^3$, the power tracking demonstrated a similar performance, with the RMS errors being close to each other. Thus, within a reasonable range, penalty factor $w$ does not significantly affect the tracking performance.

Fig. 6b shows the power-tracking curve under the proposed control strategy. To optimize the dynamic fatigue loading, penalty factor $w$ was set to a fixed value of $1\times10^3$. Load equalization was incorporated into the optimization objective by adjusting penalty factor $s$, which is the trade-off between the dynamic fatigue load and load equalization. The power-tracking error remains small, and as $s$ decreases, the power error slightly decreases. This may be because the reference power was better distributed in a different manner after the introduction of load equalization. Therefore, although a load equalization object was added, the proposed control strategy does not introduce additional power-tracking errors.

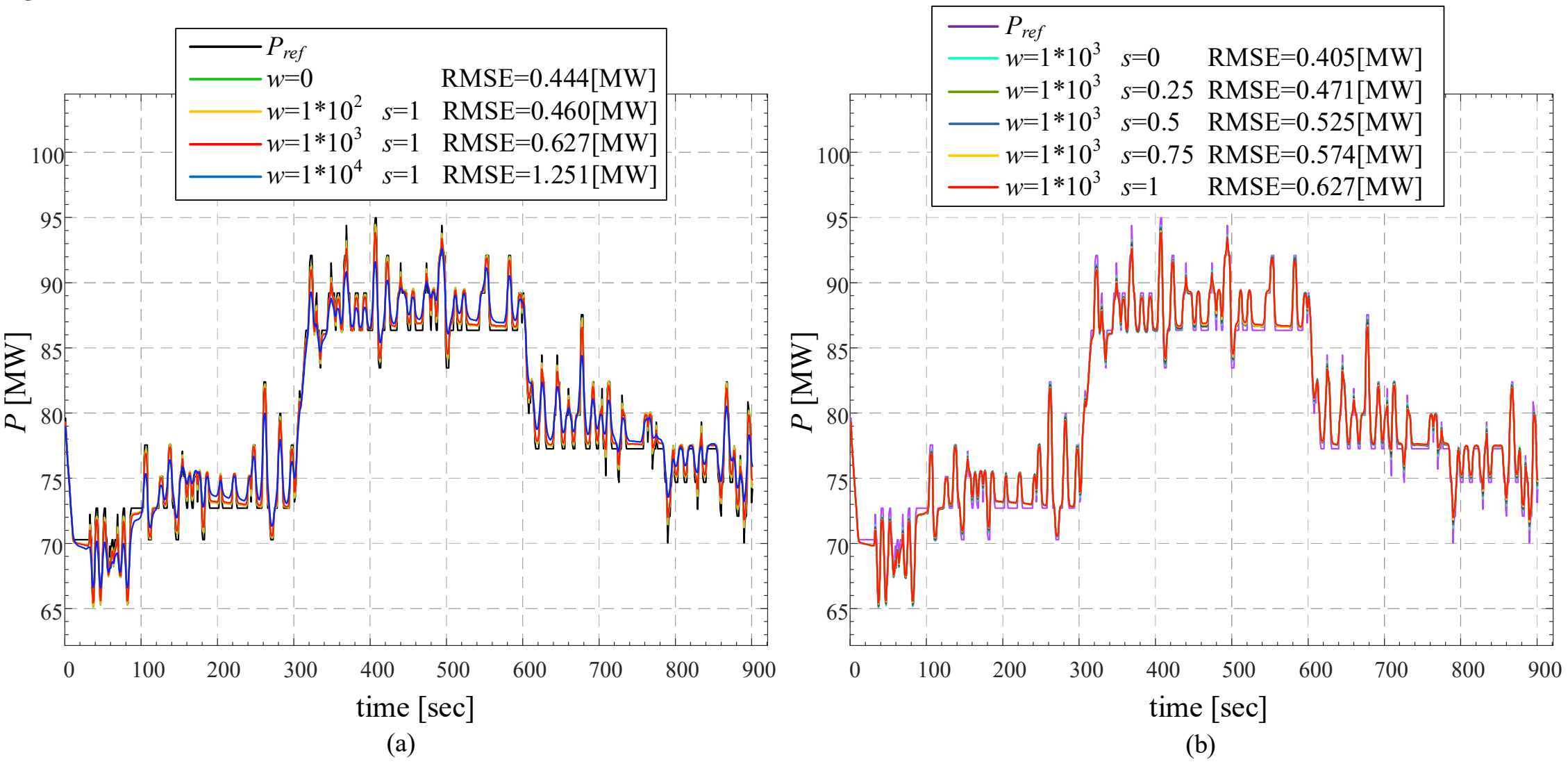

Fig. 6. Power-tracking curves with different control strategies at 9 m/s wind speed.

To evaluate the fatigue load performance of the proposed strategy, we introduce the following evaluation criteria.

$$dF_i = \sqrt{\frac{\sum_{t=1}^{900}\left(F_i(t) - F_i(t-1)\right)^2}{900}} \qquad (14)$$

$$dF = \sum_{i=1}^{64} dF_i$$

$$eF_i = \sqrt{\frac{\sum_{t=1}^{900}\left(F_i(t) - F_{mean}(t)\right)^2}{900}} \qquad (15)$$

$$eF = \sum_{i=1}^{64} eF_i$$

WT performance index $\mathrm{dF_i}$ represents the dynamic fatigue load of the $i^{th}$ turbine, whereas dF represents the dynamic fatigue load at the farm level. $\mathrm{eF_i}$ represents the deviation of the load of the $i^{th}$ turbine from the average value, and eF represents the load deviation at the farm level.

The normalized dF and eF values are shown in Fig. 7 to evaluate the fatigue load performance at the farm level. The value $w = 0$ was used for normalization, where the optimization objective only considered power tracking.

As shown in Fig. 7, when $s = 1$, the control strategy proposed in reference [20] is employed. The results indicated that as $w$ increased, the dynamic fatigue load $\overline{\mathrm{dF}}$ decreased significantly. However, there was also a certain degree of decrease in the value of $\overline{\mathrm{eF}}$. This may be attributed to the coupling relationship between the two fatigue-load optimization objectives. When $w = 1\times10^4$, there was a significant increase in the power-tracking error, which is not discussed here.

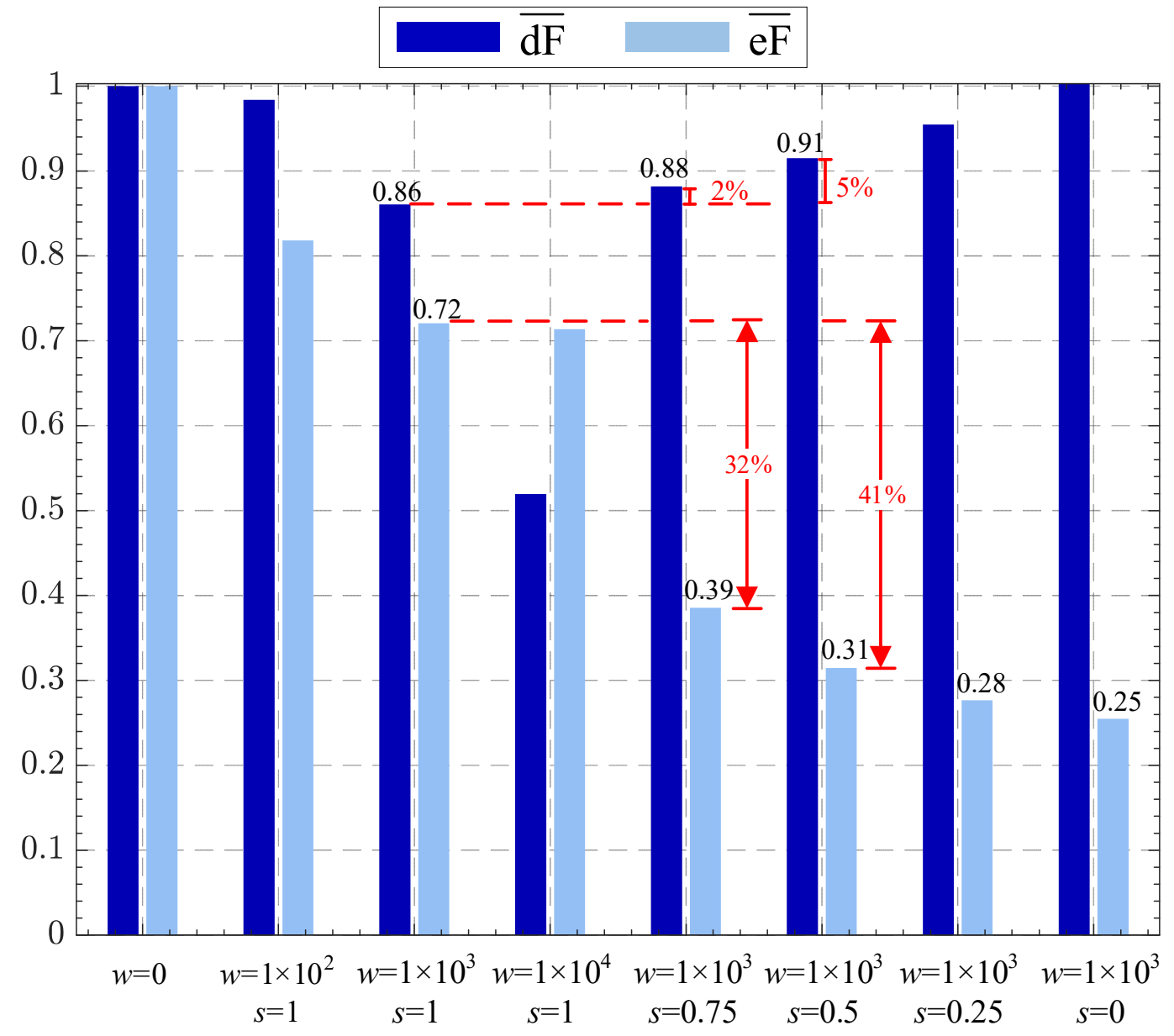

Fig. 7. Load performance on WF level under different control strategies.

The proposed control strategy incorporates load equalization into the optimization objective. To optimize the fatigue loading, $w = 1\times10^3$ and $s = 1$ were used as the baseline. When $s$ decreased from 1 to 0.75, $\overline{\mathrm{eF}}$ decreased significantly. Compared with the case of $w = 1\times10^3$ and $s = 1$, $\overline{\mathrm{eF}}$ decreased by 33%, whereas the corresponding $\overline{\mathrm{dF}}$ showed only a slight increase of 2%. As $s$ further decreased from 0.75 to 0.5, $\overline{\mathrm{eF}}$ decreased by 41%, compared to the case of $w = 1\times10^3$ and $s = 1$, whereas $\overline{\mathrm{dF}}$ increased by 5%. However, as $s$ continued to decrease, $\overline{\mathrm{dF}}$ continued to increase and $\overline{\mathrm{eF}}$ decreased insignificantly.

In summary, by setting the penalty coefficient $s$ in the range 0.5–0.75, a significant improvement in load equalization optimization can be achieved. In this setting, $\overline{eF}$ decreases by more than 33%, whereas $\overline{dF}$ increases by less than 5%. This suggests that, within a reasonable range of penalty factors, the controller can effectively achieve load equalization optimization, reduce load variations among WTs, and maintain the impact on dynamic fatigue loads within a lower range. This provides a reliable and durable solution for the operation and maintenance of WFs.

Because the wind direction was set to 0°, the control results for the WTs located in the same row were almost identical. To illustrate the results, $WT_{25\text{-}32}$ was selected as the representative turbine in the WF. In Fig. 4, the turbines are highlighted with dashed black lines representing the upstream, midstream, and downstream positions within the WF.

Fig. 8 illustrates the fatigue load distribution of a single turbine under the control strategy reported in [20]. As shown in Fig. 8a, when penalty factor $w = 0$, the upstream turbines ($WT_{25}$ and $WT_{26}$) exhibit the highest $dF_i$ values. As $w$ increased to $1\times10^3$, as represented by the red bars in Fig. 8a, the $dF_{25,26}$ values significantly decreased, while the $dF_{27\text{-}29}$ values remained relatively stable, and the $dF_{30\text{-}32}$ values slightly increased. When $w = 1\times10^4$, all $dF_i$ values decreased. However, as mentioned earlier, there was a significant increase in the power-tracking error, which is not discussed here. As shown in Fig. 8b, as $w$ increased, $eF_i$ decreased. This indicates a certain coupling between the two load optimization objectives.

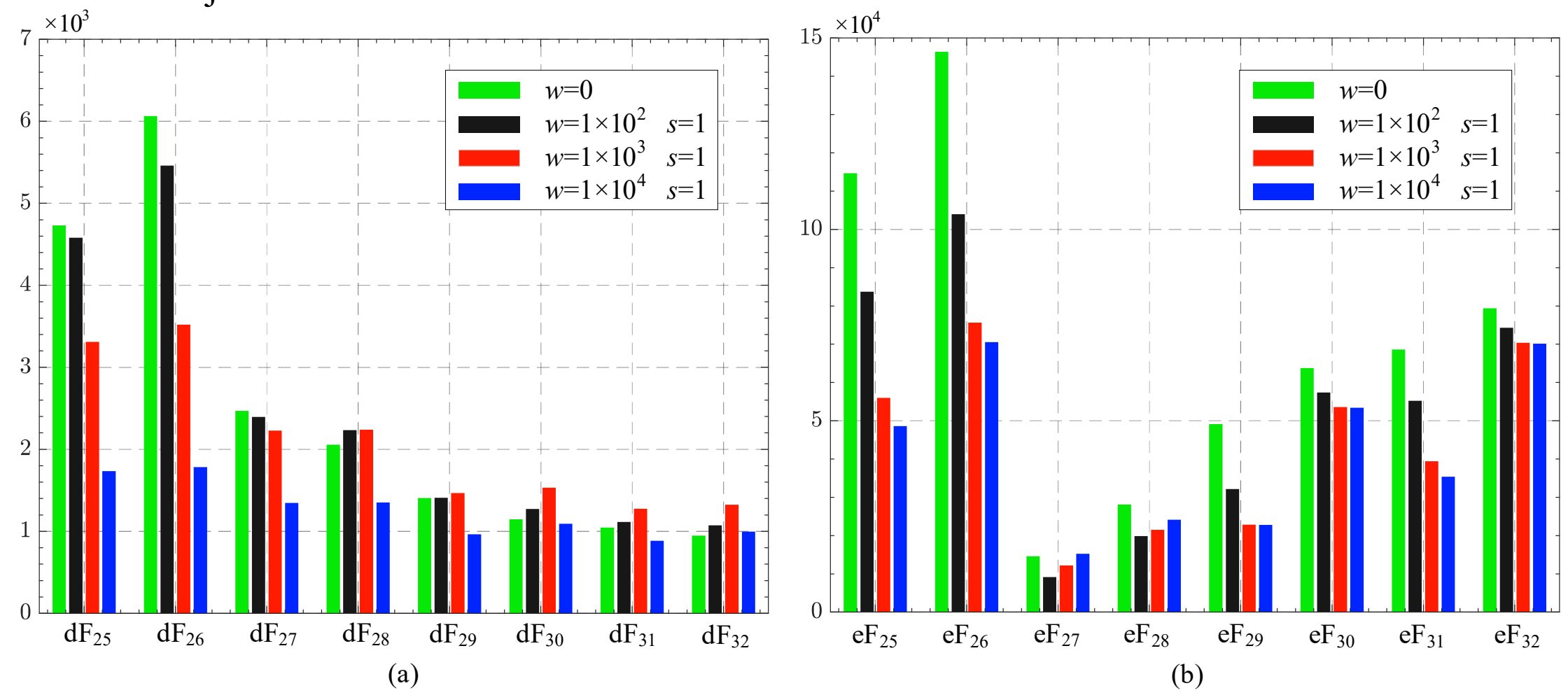

Fig. 8. Load performance using the control strategy reported in reference [20].

Fig. 9 shows the fatigue load performance obtained when the proposed control strategy was used. To optimize the fatigue load distribution, $w = 1\times10^3$ and $s = 1$ were used as the baselines, which correspond to the red bars in Fig. 8. In Fig. 9b, when $s$ decreases from 1 to 0.75, load equalization performance index $eF_i$ significantly decreases for all WTs. Simultaneously, slightly increases as shown in in Fig. 9a, $dF_{25,26}$. However, when $s$ further decreases from 0.75 to 0, the improvement in load equalization optimization becomes less significant and $dF_{25,26}$ significantly increases.

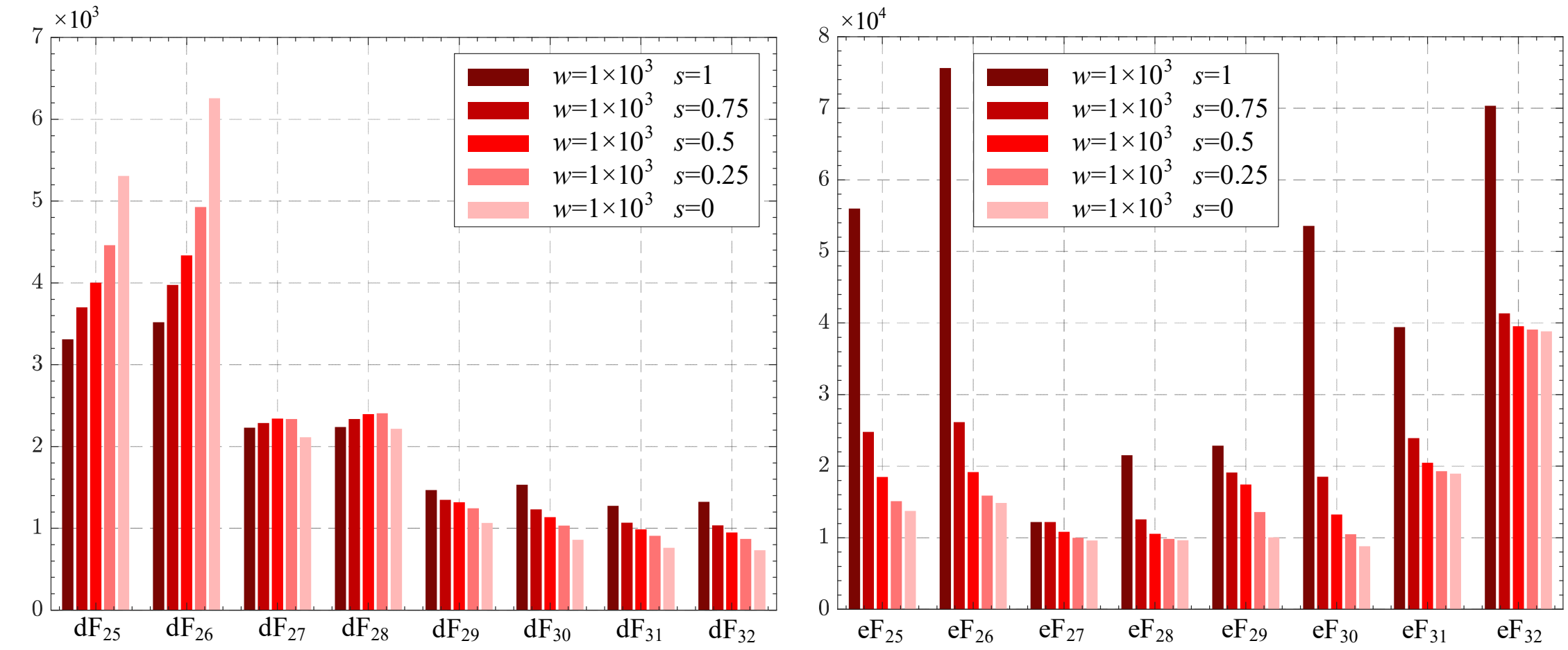

(a) (b)

Fig. 9. Load performance obtained using the proposed control strategy.

*4.2. Average wind speed of 12 m/s.*

The power-tracking curves for two control algorithms at an average wind speed of 12 m/s are shown in Fig. 10. Fig. 10a shows the power-tracking curves obtained using the method from [2] with different $w$ values. When $w = 1\times10^4$ and $s = 1$, the power-tracking error was large. The remaining groups of penalty factors exhibited similar tracking performances with acceptable errors.

Fig. 10b, shows the power-tracking curve obtained using the proposed method. When $w$ was set to a fixed value of $1\times10^3$, the power-tracking error decreased slightly as $s$ decreased, which is consistent with the scenario with a wind speed of 9 m/s.

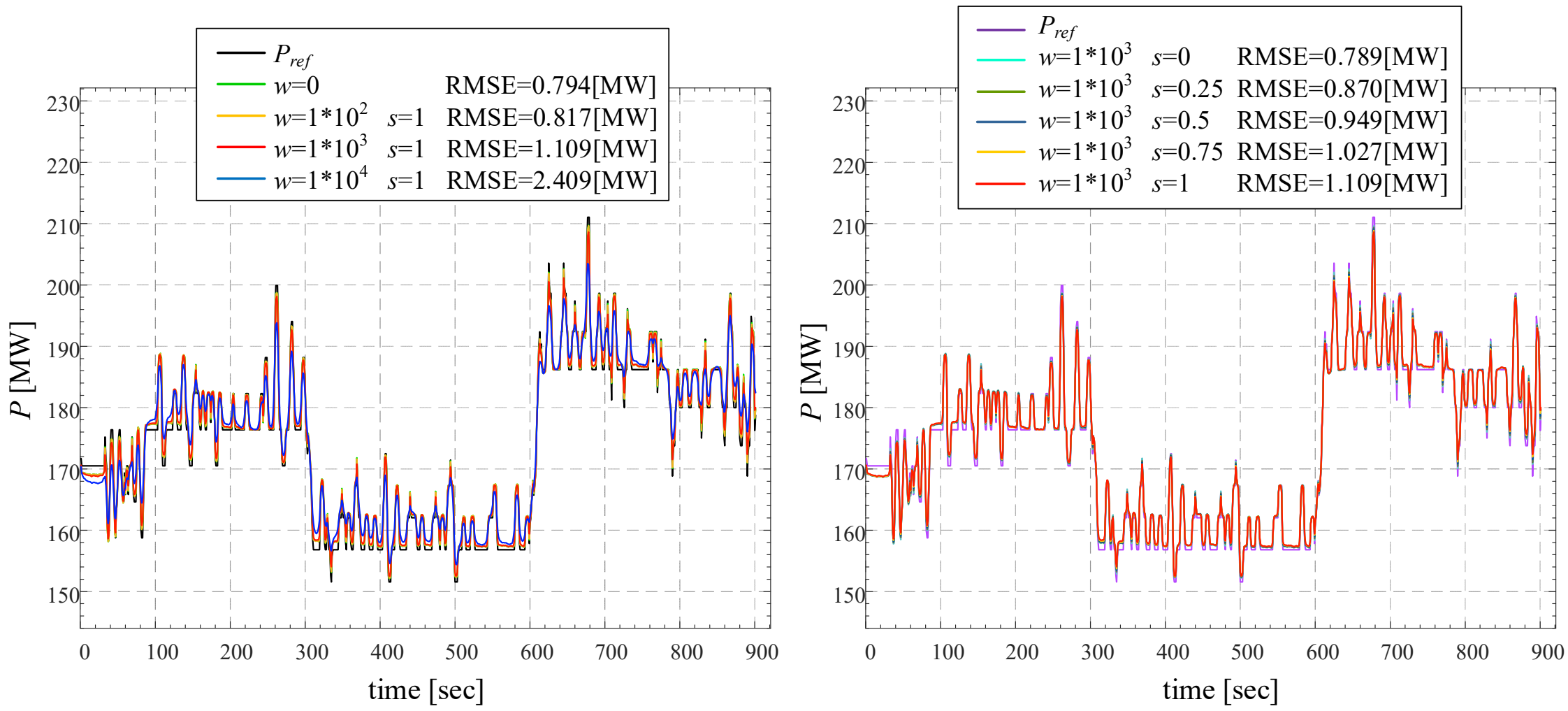

(a) (b)

Fig. 10. Power-tracking curves with different control strategies at a wind speed of 12 m/s.

The normalized dF and eF values are shown in Fig. 11. In the case of $s = 1$, the method from [20] was used. As $w$ increases, $\overline{\mathrm{dF}}$ decreases significantly, but there is also a certain decrease in $\overline{\mathrm{eF}}$, consistent with the scenario with a wind speed of 9 m/s. When $w = 1\times10^4$, the power-tracking error is large, which is not discussed here. $w = 1\times10^3$ and $s = 1$ were used for comparison. When $s$ decreased to 0.75 and 0.5, $\overline{eF}$ decreased significantly. Compared to the cases of $w = 1\times10^3$ and $s = 1$, $\overline{eF}$ decreased by 39% and 50%, respectively, whereas the corresponding $\overline{dF}$ showed only a slight increase, with increases of 2% and 5%, respectively. As $s$ decreased, $\overline{dF}$ increased; however, $\overline{eF}$ decreased insignificantly.

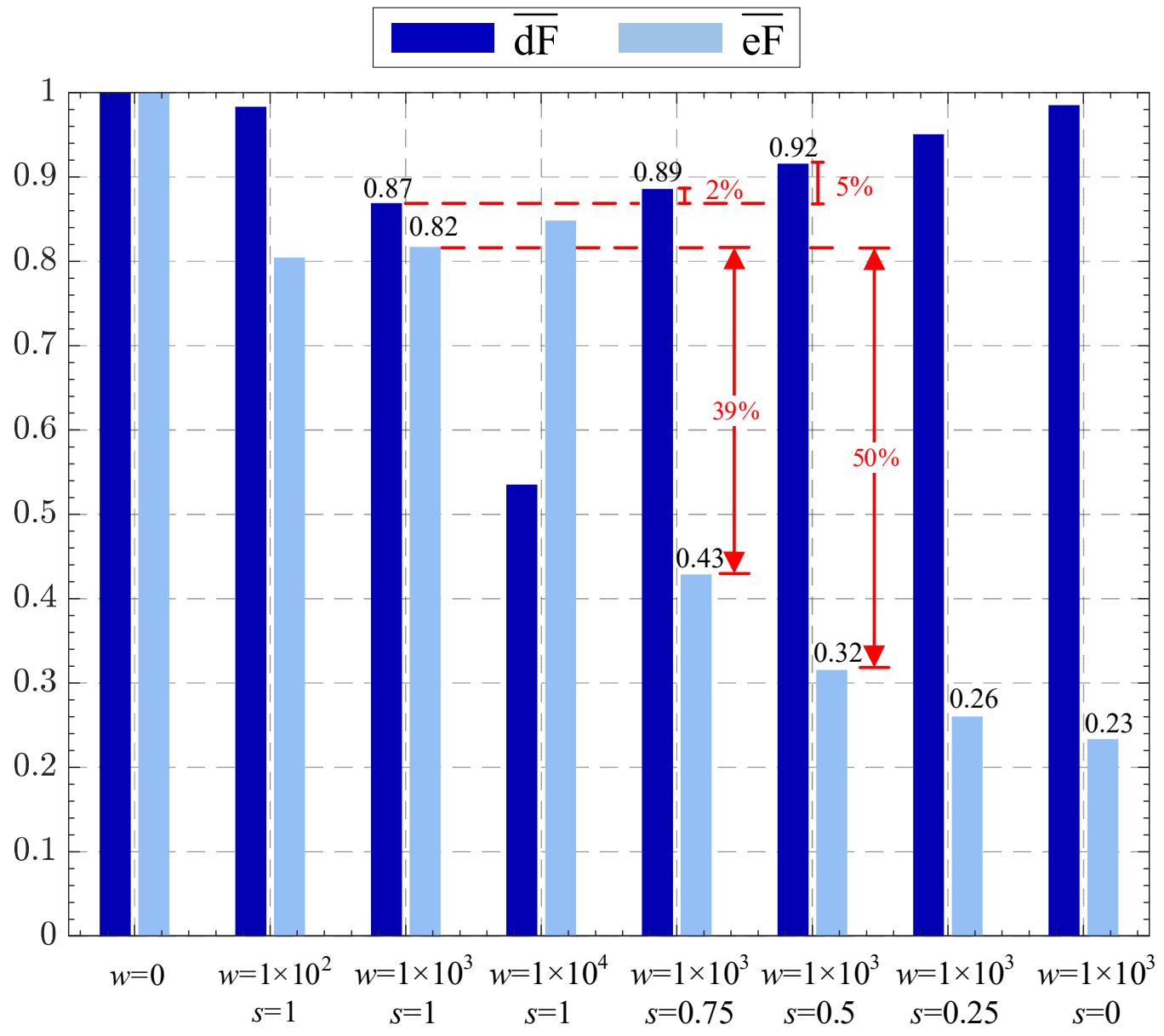

Fig. 11. Load performance on WF level for different control strategies.

Fig. 12 shows the fatigue load distribution results obtained using the method described in [20]. As shown in Fig. 12a, as $w$ increases, the $dF_{25,26}$ values significantly decrease while the $dF_{27\text{-}29}$ values remain relatively stable and $dF_{30\text{-}32}$ values increase slightly. When $w = 1\times10^4$, all $dF_i$ values decrease, and this significantly affects the power-tracking performance; therefore, this scenario is not considered. As shown in Fig. 12b, when $w = 0$, $eF_{25\text{-}26}$ is high. As $w$ increases, $eF_i$ decreases to a certain extent, which is consistent with the case of a wind speed of 9 m/s.

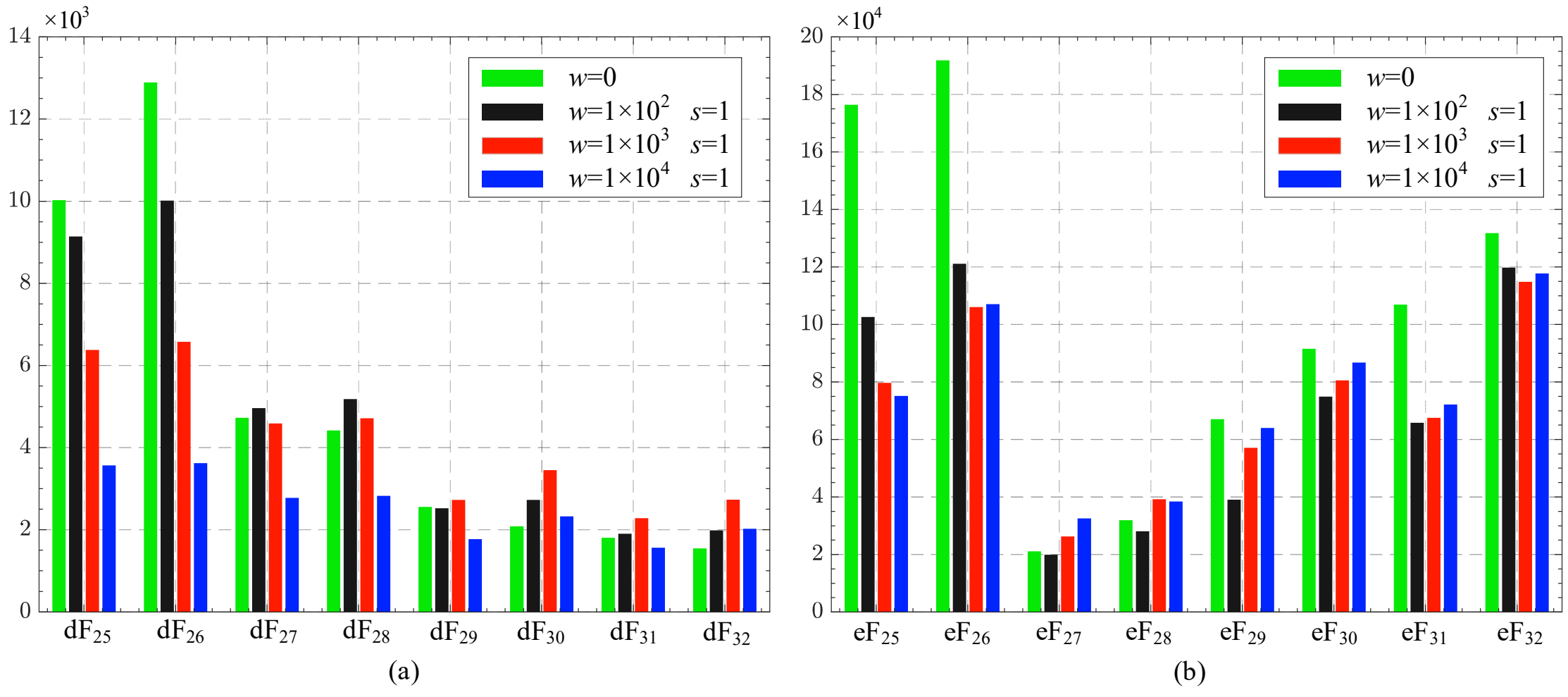

Fig. 12. Load performance using the control strategy reported in reference [20].

Fig. 13 shows the distribution results obtained when the proposed control strategy was used. $w = 1\times10^3$ and $s = 1$ (red bars in Fig. 12) was used as the reference comparison. In the figure, when $s$ decreases from 1 to 0.75, all the $eF_i$ values decrease significantly, indicating a significant improvement in the load equalization optimization. As $s$ further decreases from 0.75 to 0.5, some turbines show a noticeable decrease in $eF_i$, accompanied by a slight increase in $dF_i$. Compared to the low wind speed case, the controller achieves better load equalization optimization at an average wind speed of 12 m/s. However, when $s$ decreases from 0.5 to 0, the load equalization optimization effect becomes less apparent, whereas $dF_{25\text{-}26}$ increases significantly.

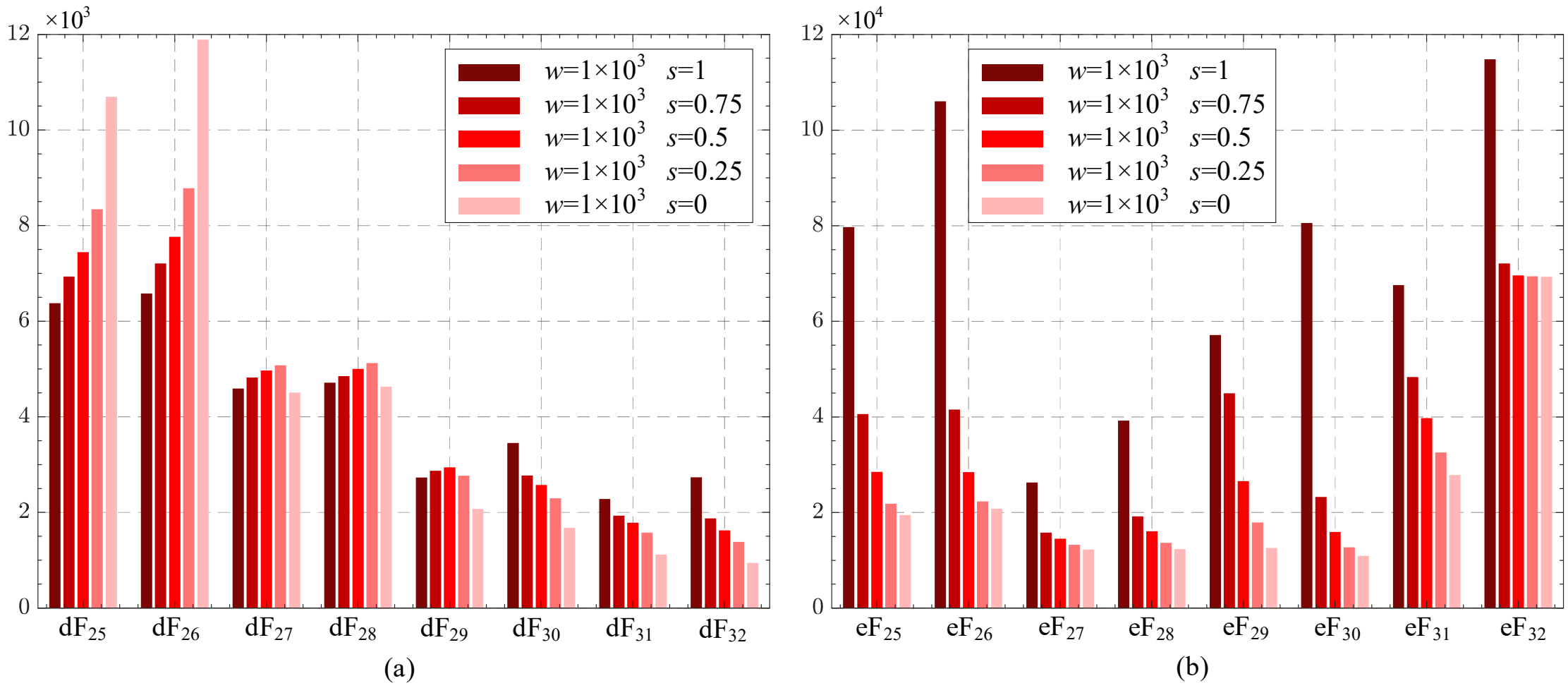

Fig. 13. Load performance using the proposed control strategy.

## 5. Conclusion

In this study, we proposed a closed-loop model predictive controller for WF participation in frequency regulation. The power distribution among the WTs was formulated as a constrained optimization problem to simultaneously minimize the power-tracking error, dynamic fatigue load, and load deviation from the average value (load equalization). The proposed strategy was tested in WFsim, a medium-fidelity model. Comparing with the recently proposed control strategies, the following conclusions were obtained: 1) In WFs, the fatigue load can be reduced while having no significant impact on the power-tracking performance. 2) Load equalization was incorporated into the optimization objective by adjusting the weightings of the dynamic fatigue load and load equalization using penalty factor $s$. When penalty factor $s$ was within the range 0.5–0.75, significant improvements in load equalization were observed. For wind speeds of 9 and 12 m/s, the load-equalization index (eF) in the WF decreased by more than 33% and 39%, respectively, whereas the dynamic fatigue load index (dF) increased by less than 5%. Therefore, within a reasonable range of penalty factors, the controller can effectively achieve load equalization optimization while maintaining the impact on the dynamic fatigue load within a lower range.

In future work, the wake dynamic characteristics must be analyzed in large-scale WFs, as this can lead to more efficient control performance.

## Appendix A

*Nomenclature:*

WT: Wind turbine
WF: Wind farm
MPC: Model predictive control
$\rho$: Air density

$U$: Wind speed
$P_{ref}$: Reference power
$P_{command}$: Power command
$f_{measure}$: Measured frequency
$K_D$: Droop coefficient
$K_I$: Virtual inertia coefficient
$F$: Axial force
$P$: Power output
$C_T'$ : Local thrust coefficient
$\hat{C}_{T_i}'(t)$ : first-order filter signals
$\tau$: Time constant of the filter
$N$: Number of wind turbines
$M$: Prediction period
D: Rotor diameter
$\sigma$: Turbulence intensity

*Design Values of Considered System Parameters:*

$\rho = 1.2$ kg/m$^3$, $K_D = 50$, $K_I = 10$, $\tau = 5$ s, D = 126 m, $\sigma = 0.1$, $M = 10$ s, $N = 64$, $C'_{T\min} = 0.1$, $C'_{T\max} = 2$, $dC'_T = 0.2$; the simulation sample time was 1 s, the constant value $r$ in the weight matrix $R$ was set to 1e12.

## Acknowledgements

This work was supported by the National Natural Science Foundation of China (Grant No. 52077030), the Jilin Province Science and Technology Development Plan Project (Grant No. 20190201289JC)

## References

[1] Díaz, Hugo, and C. Guedes Soares. "Review of the current status, technology and future trends of offshore wind farms." Ocean Engineering 209 (2020): 107381.
[2] Ahmad, Tanvir, et al. "Fast processing intelligent wind farm controller for production maximisation." Energies 12.3 (2019): 544.
[3] Gebraad, Pieter MO, and Jan-Willem van Wingerden. "Maximum power-point tracking control for wind farms." Wind Energy 18.3 (2015): 429-447.
[4] Park, Jinkyoo, and Kincho H. Law. "A data-driven, cooperative wind farm control to maximize the total power production." Applied Energy 165 (2016): 151-165.
[5] Bossuyt J, Howland M F, Meneveau C, et al. Measurement of unsteady loading and power output variability in a micro wind farm model in a wind tunnel[J]. Experiments in Fluids, 2017, 58(1): 1-17.
[6] Menezes E J N, Araújo A M, Da Silva N S B. A review on wind turbine control and its associated methods[J]. Journal of cleaner production, 2018, 174: 945-953.
[7] Fu Y, Wang Y, Zhang X. Integrated wind turbine controller with virtual inertia and primary frequency responses for grid dynamic frequency support[J]. IET Renewable Power Generation, 2017, 11(8): 1129-1137.
[8] Van de Vyver J, De Kooning J D M, Meersman B, et al. Droop control as an alternative inertial response strategy for the synthetic inertia on wind turbines[J]. IEEE Transactions on Power Systems, 2015, 31(2): 1129-1138.
[9] Barik, Amar Kumar, and Dulal Chandra Das. "Optimal load-frequency regulation of BioRenewable cogeneration based interconnected hybrid microgrids with demand response support." 2018 15th IEEE India Council International Conference (INDICON). IEEE, 2018.
[10] Liao, Hao, et al. "Active power dispatch optimization for offshore wind farms considering fatigue distribution." Renewable energy 151 (2020): 1173-1185.
[11] Bay C J, Annoni J, Taylor T, et al. Active power control for wind farms using distributed model predictive control and nearest neighbor communication[C]//2018 Annual American Control Conference (ACC). IEEE, 2018: 682-687.

[12] Chen S, Mathieu J L, Seiler P. Stochastic model predictive controller for wind farm frequency regulation in waked conditions[J]. Electric Power Systems Research, 2022, 211: 108543.

[13] Boersma S, Doekemeijer B M, Keviczky T, et al. Stochastic model predictive control: uncertainty impact on wind farm power tracking[C]//2019 American Control Conference (ACC). IEEE, 2019: 4167-4172.

[14] Boersma S, Rostampour V, Doekemeijer B, et al. A constrained model predictive wind farm controller providing active power control: an LES study[C]//Journal of Physics: Conference Series. IOP Publishing, 2018, 1037(3): 032023.

[15] Vali, Mehdi, et al. "Model predictive active power control of waked wind farms." 2018 Annual American Control Conference (ACC). IEEE, 2018.

[16] Buccafusca, Lucas, and Carolyn Beck. "Multiobjective model predictive control design for wind turbines and farms." Journal of Renewable and Sustainable Energy 13.3 (2021): 033312.

[17] Howland, Michael F., et al. "Collective wind farm operation based on a predictive model increases utility-scale energy production." Nature Energy 7.9 (2022): 818-827.

[18] Zhao H, Wu Q, Guo Q, et al. Distributed model predictive control of a wind farm for optimal active power controlpart I: Clustering-based wind turbine model linearization[J]. IEEE transactions on sustainable energy, 2015, 6(3): 831-839.

[19] Riverso S, Mancini S, Sarzo F, et al. Model predictive controllers for reduction of mechanical fatigue in wind farms[J]. IEEE Transactions on Control Systems Technology, 2016, 25(2): 535-549.

[20] Boersma S, Doekemeijer B M, Siniscalchi-Minna S, et al. A constrained wind farm controller providing secondary frequency regulation: An LES study[J]. Renewable energy, 2019, 134: 639-652.

[21] Vali M, Petrović V, Pao L Y, et al. Lifetime extension of waked wind farms using active power control[C]//Journal of Physics: Conference Series. IOP Publishing, 2019, 1256(1): 012029.

[22] Vali M, Petrović V, Steinfeld G, et al. An active power control approach for wake-induced load alleviation in a fully developed wind farm boundary layer[J]. Wind Energy Science, 2019, 4(1): 139-161.

[23] Vali M, Petrović V, Pao L Y, et al. Model predictive active power control for optimal structural load equalization in waked wind farms[J]. IEEE Transactions on Control Systems Technology, 2021, 30(1): 30-44.

[24] Boersma S, Doekemeijer B, Vali M, et al. A control-oriented dynamic wind farm model: WFSim[J]. Wind Energy Science, 2018, 3(1): 75-95.

[25] Versteeg, Henk Kaarle, and Weeratunge Malalasekera. An introduction to computational fluid dynamics: the finite volume method. Pearson education, 2007.

[26] Ochoa D, Martinez S. Fast-frequency response provided by DFIG-wind turbines and its impact on the grid[J]. IEEE Transactions on Power Systems, 2016, 32(5): 4002-4011.

[27] González-Longatt, Francisco, P. Wall, and V. Terzija. "Wake effect in wind farm performance: Steady-state and dynamic behavior." Renewable Energy 39.1 (2012): 329-338.

[28] Wang, Zhongguan, and Wenchuan Wu. "Coordinated control method for DFIG-based wind farm to provide primary frequency regulation service." IEEE Transactions on Power Systems 33.3 (2017): 2644-2659.

[29] Yao, Qi, et al. "New design of a wind farm frequency control considering output uncertainty and fatigue suppression." Energy Reports 9 (2023): 1436-1446.

[30] Vali, Mehdi, et al. "Adjoint-based model predictive control for optimal energy extraction in waked wind farms." Control Engineering Practice 84 (2019): 48-62.

[31] Silva J G, Ferrari R, van Wingerden J W. Wind farm control for wake-loss compensation, thrust balancing and load-limiting of turbines[J]. Renewable Energy, 2023, 203: 421-433.

[32] González, Héctor Del Pozo, and José Luis Domínguez-García. "Non-centralized hierarchical model predictive control strategy of floating offshore wind farms for fatigue load reduction." Renewable Energy 187 (2022): 248-256.

[33] Ke, Shitang, et al. "Wind‐induced fatigue of large HAWT coupled tower–blade structures considering aeroelastic and yaw effects." The Structural Design of Tall and Special Buildings 27.9 (2018): e1467.